\documentclass[11pt]{article}
\usepackage{xspace}
\usepackage{ifpdf}
\usepackage{graphicx_color}
\usepackage{moriond,epsfig}

\def\Journal#1#2#3#4{{#1} {\bf #2}, #3 (#4)}

\def\PLB{{\em Phys. Lett.}  B}
\def\PRL{\em Phys. Rev. Lett.}
\def\PRD{{\em Phys. Rev.} D}

\def\be{\begin{equation}}
\def\ee{\end{equation}}
\def\bea{\begin{eqnarray}}
\def\eea{\end{eqnarray}}

\def\fns{\footnotesize}
\def\ppb{\mbox{$p\overline{p}$}\xspace}
\def\invfb{\mbox{fb$^{-1}$}\xspace}
\begin{document}

\begin{flushright}
FERMILAB-CONF-12-200-PPD \\
\end{flushright}

\vspace*{3.6cm}
\title{NEW MEASUREMENTS WITH PHOTONS AT THE TEVATRON}

\author{ J.R. DITTMANN \\ (on behalf of the CDF and D0 collaborations) }

\address{Department of Physics, Baylor University, \\ One Bear Place \#97316,
Waco TX 76798-7316, USA}

\maketitle\abstracts{
We present three recent photon analyses from data collected at the Fermilab Tevatron:  measurements of the direct photon pair production cross section at CDF and D0, measurements of azimuthal decorrelations and multiple parton interactions in $\gamma$ + 2 jet and $\gamma$ + 3 jet events at D0, and an observation of exclusive diphoton production at CDF.}

\section{Introduction}

With the recent completion of Run II at the Fermilab Tevatron, the CDF and D0 experiments are publishing results based on challenging measurements that probe quantum chromodynamics (QCD) and are sensitive to next-to-leading-order (NLO) and next-to-next-to-leading-order (NNLO) effects and non-perturbative physics.  A superior understanding of parton distribution functions and QCD backgrounds will  improve the sensitivity of searches for new phenomena at the LHC and reduce uncertainties in a multitude of future measurements.

\section{Prompt Diphoton Production at CDF and D0}

Precise measurements of the diphoton production cross section are important as a test of perturbative QCD and soft gluon resummation.  Furthermore, the production of prompt photon pairs in hadron collisions is a large background in many ongoing searches including low-mass Higgs decays to diphotons, new heavy resonances, extra spatial dimensions, and cascade decays of heavy new particles.  The measurement of prompt photon pair production at $\sqrt{s}$ = 1.96~TeV was performed by CDF using 5.36~\invfb of data and by D0 using 4.2~\invfb of data.

Prompt photons are produced directly from the hard scattering or fragmentation process as opposed to photons from the decay of particles such as $\pi^0$, $\eta$, or $K^0_s$.  At a much smaller rate ($<$ 1\%), photon pairs may come from Higgs boson decay, graviton decay (extra dimensions), or neutralino decay (SUSY).  A variety of theoretical predictions are available (e.g. \textsc{pythia}, \textsc{diphox}, and \textsc{resbos}), where each includes a different set of Feynman diagrams in the calculation.\cite{dipho}

The CDF \cite{cdfpp} and D0 \cite{dzpp} analyses both identify two isolated, high $E_T$ ($p_T$) photons in the central region.  Diphotons are identified with a purity of about 70\% among backgrounds consisting mainly of $\gamma$ + jet, dijet, and $Z/\gamma^* \to e^+ e^-$ production.  Whereas the CDF diphoton selection is cut-based, the D0 analysis uses a neural net discriminant to separate jets and photons.

\begin{figure}[t]
\begin{center}
\includegraphics[width=1.8in]{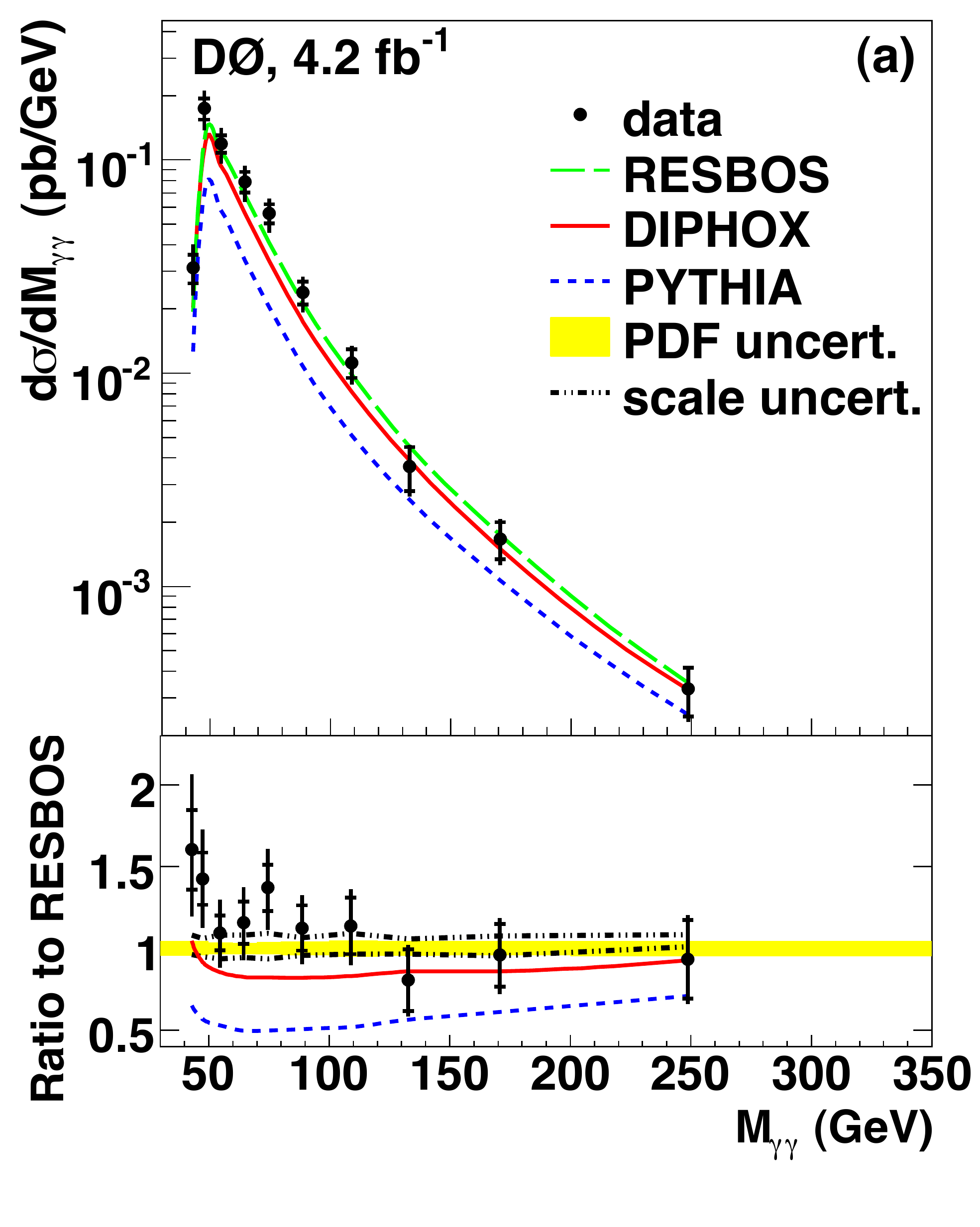}
\includegraphics[width=1.8in]{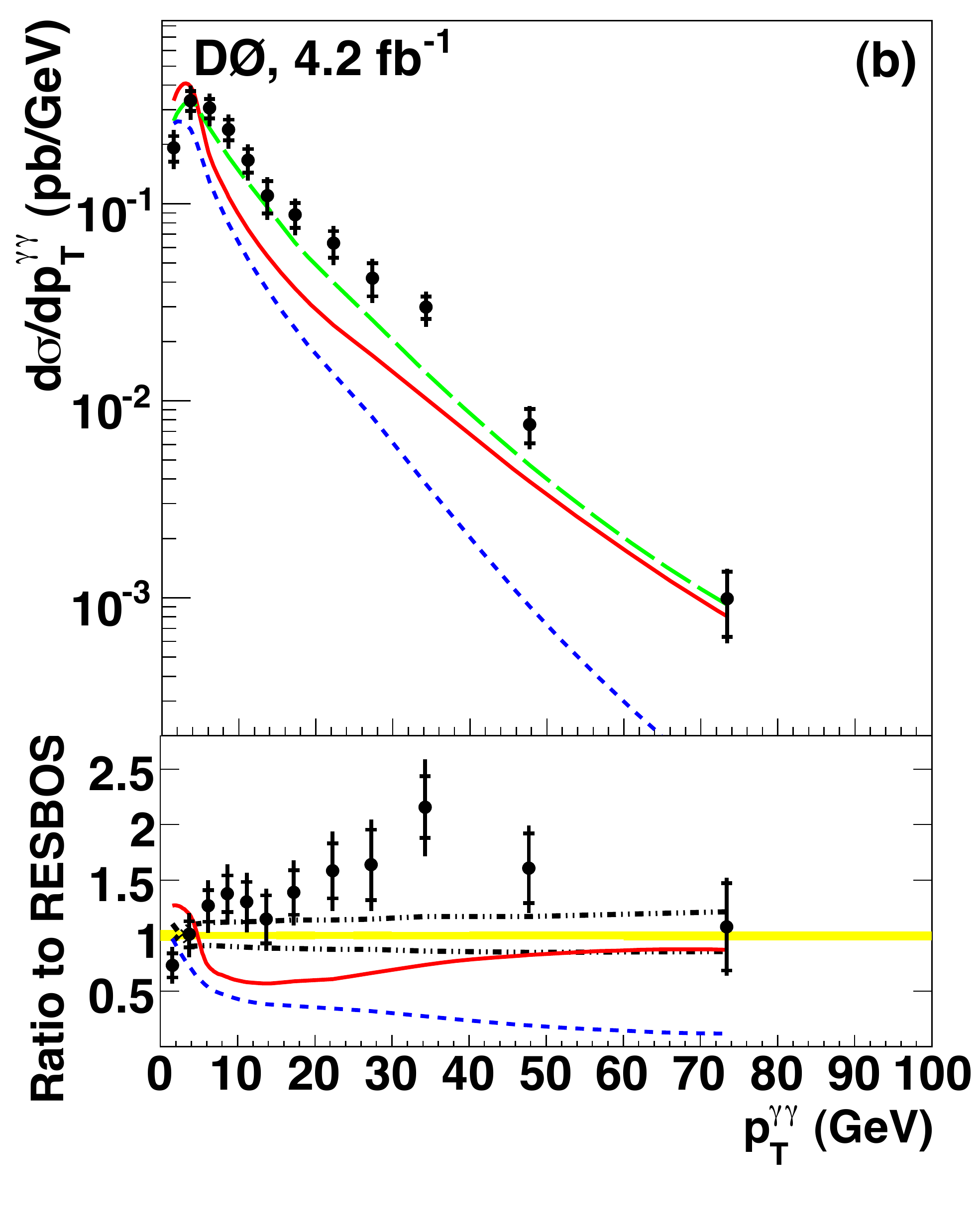}
\includegraphics[width=1.8in]{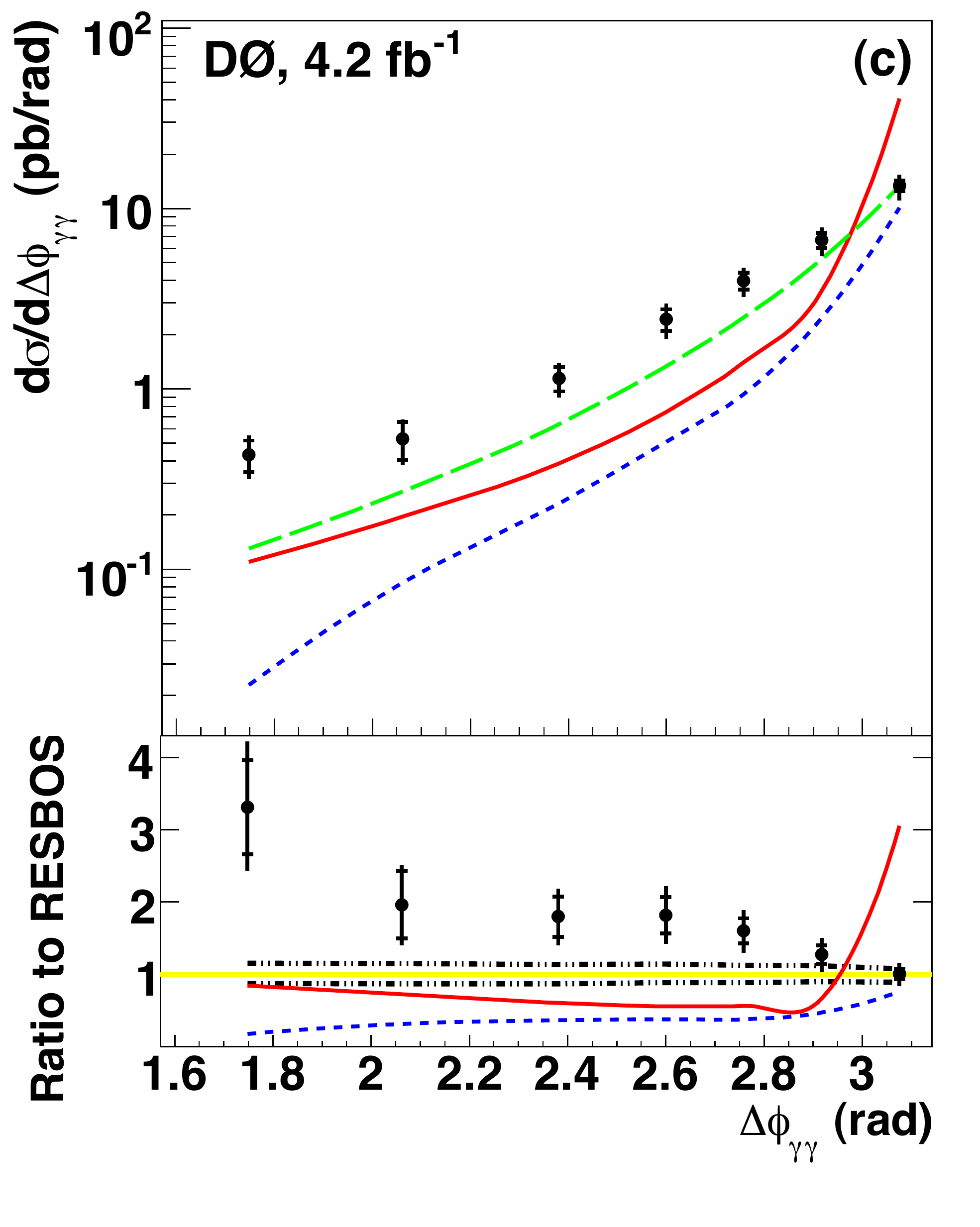}
\end{center}
\vspace*{-3ex}
\caption{The measured differential diphoton production cross sections at D0 as a function of (a) $M_{\gamma\gamma}$, (b) $p_T^{\gamma\gamma}$, and (c) $\Delta\phi_{\gamma\gamma}$.  The data are compared to theoretical predictions from \textsc{resbos}, \textsc{diphox}, and \textsc{pythia}.  
\label{fig:d0}}
\end{figure}

\begin{figure}[t!]
\begin{center}
\includegraphics[width=1.6in]{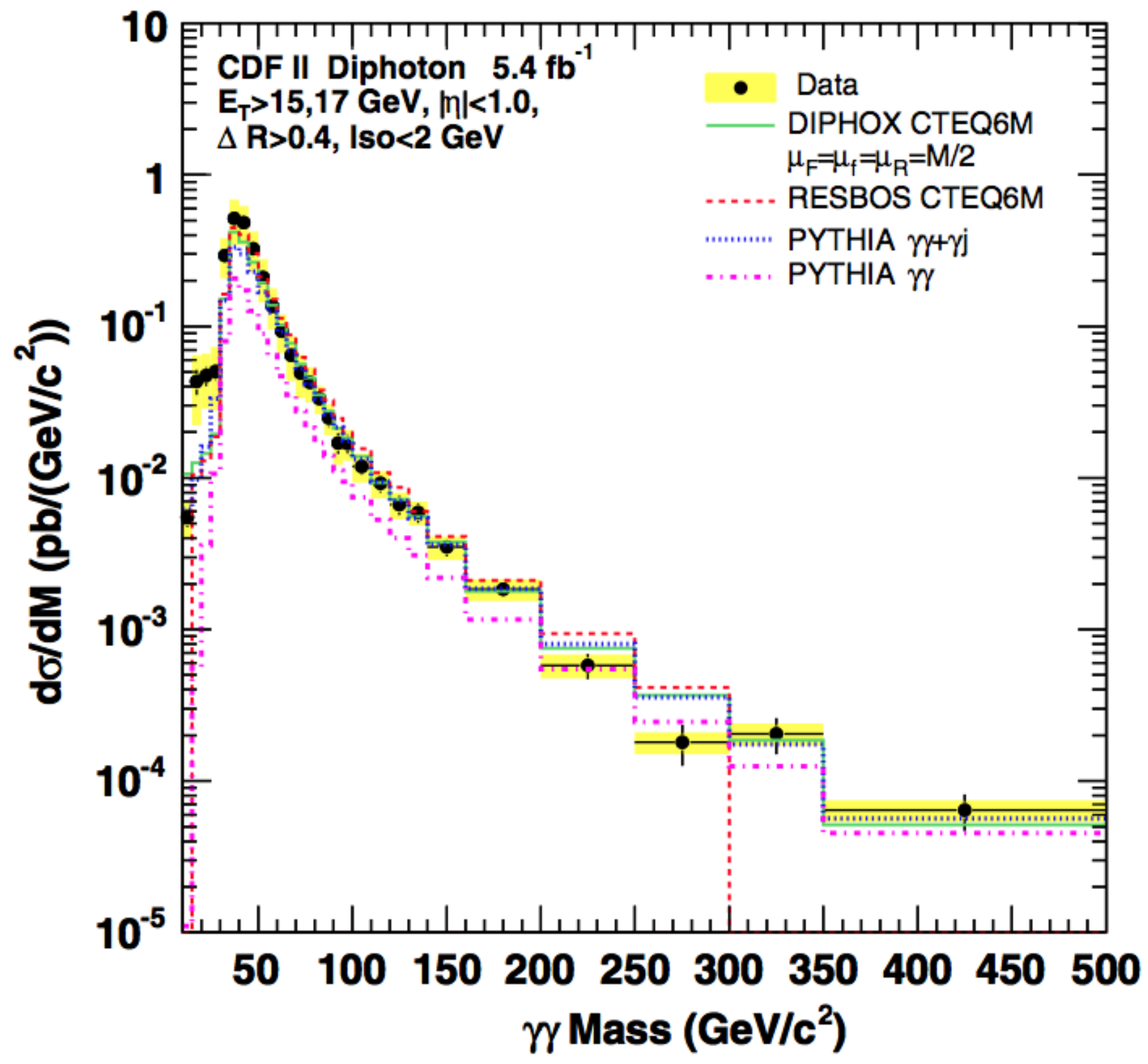}\hspace{0.05cm}
\includegraphics[width=1.6in]{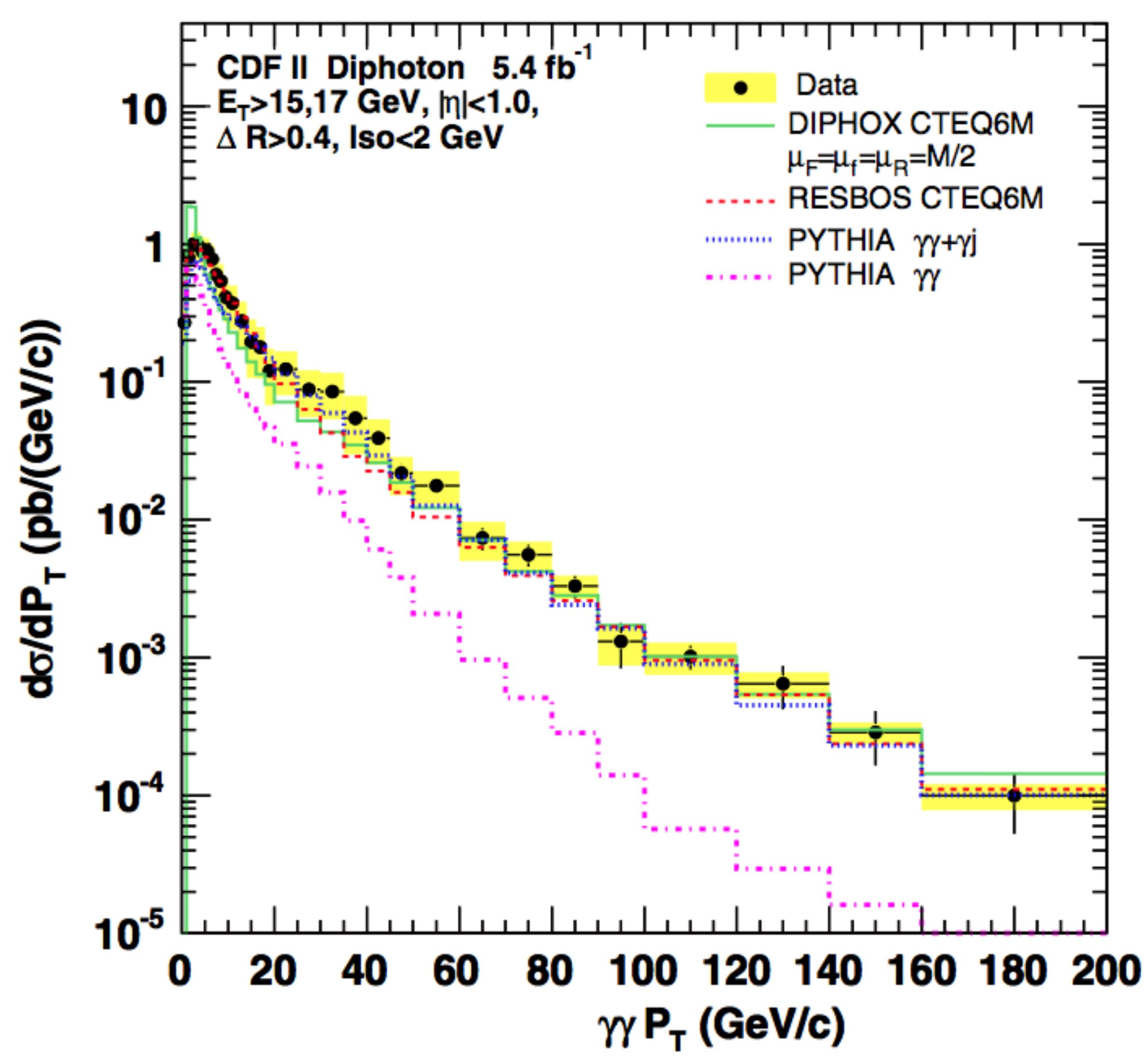}\hspace{0.1cm}
\includegraphics[width=1.6in]{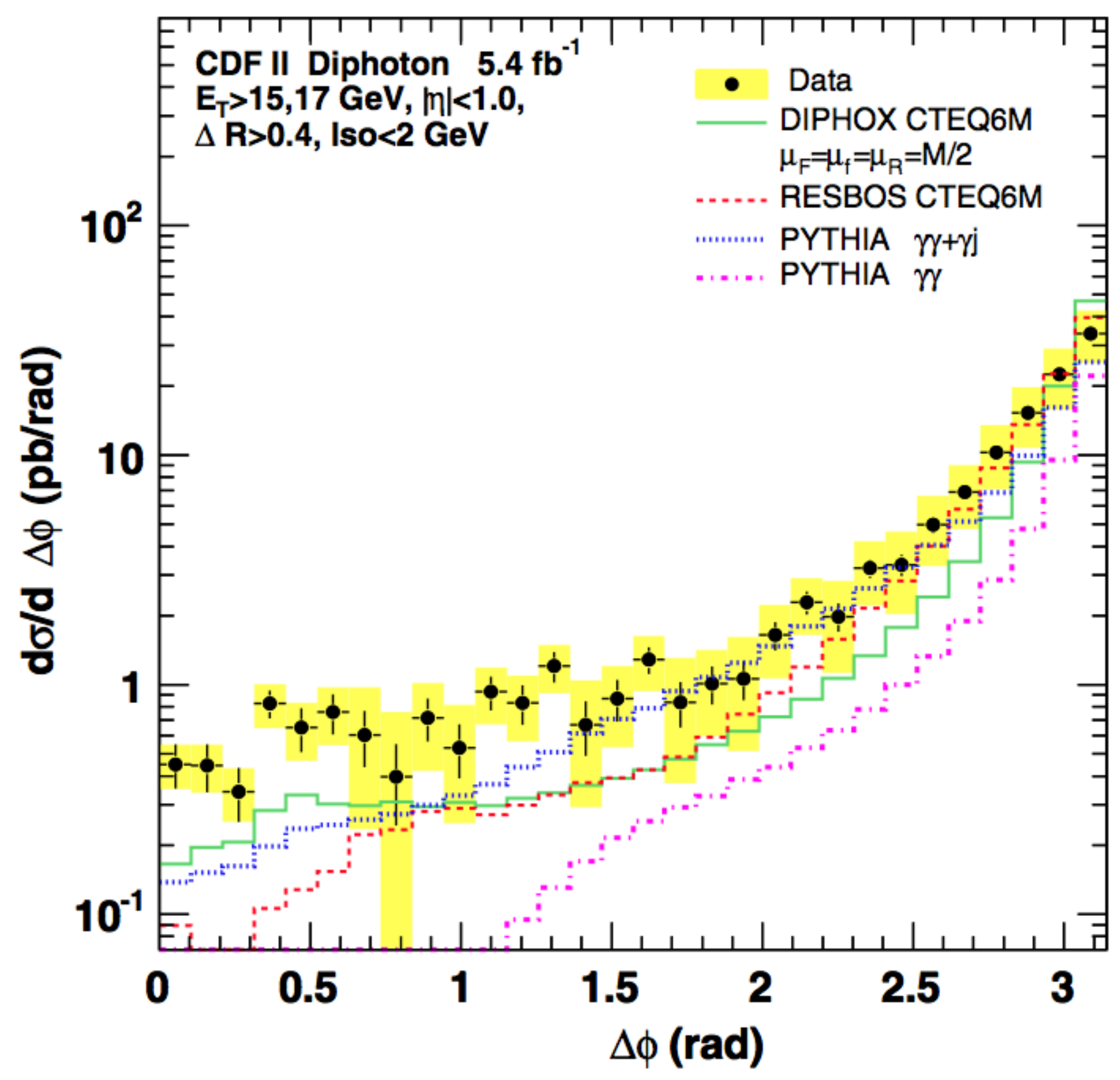}\hspace*{0.2cm} \\
\includegraphics[width=1.6in]{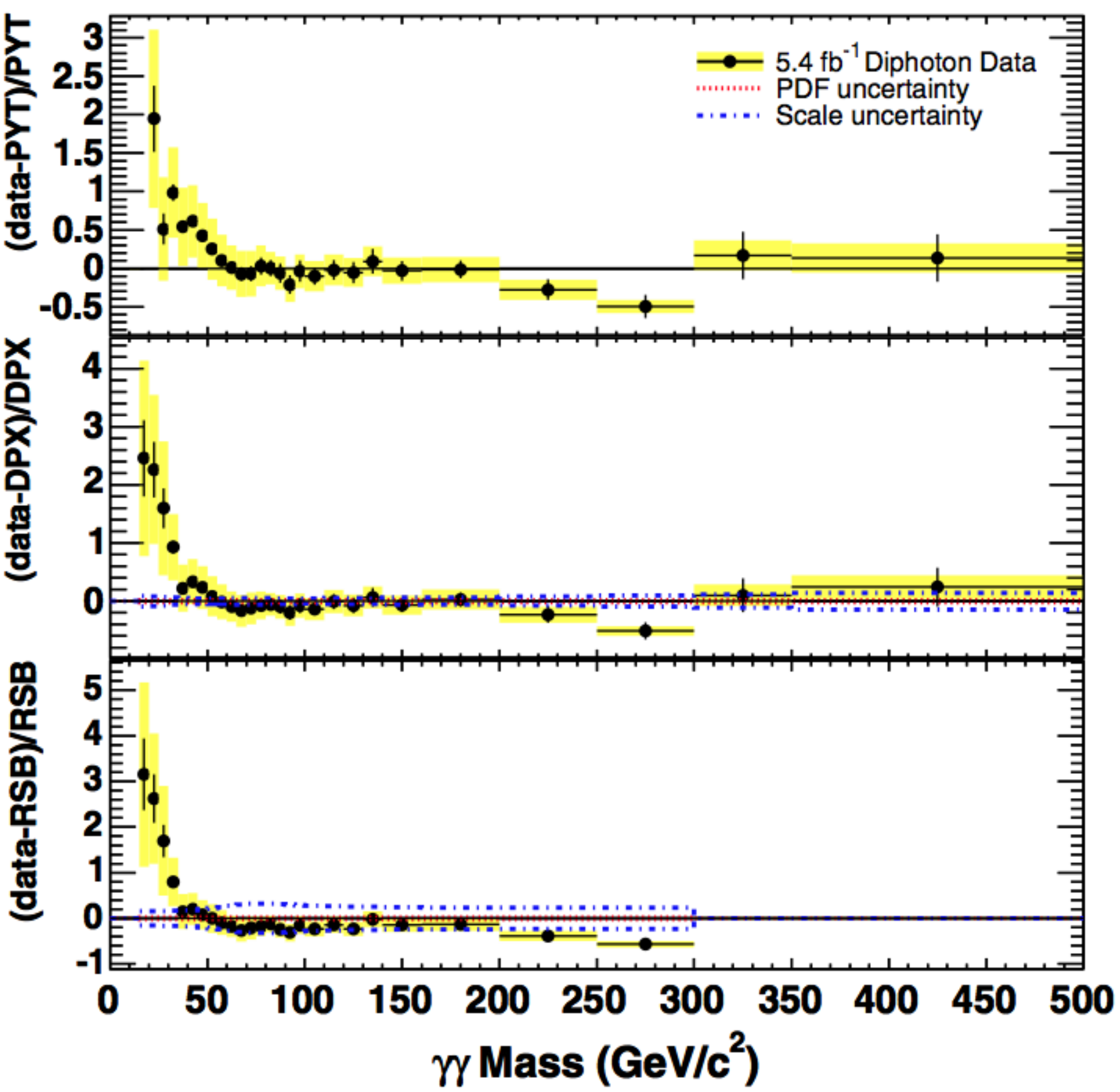}
\includegraphics[width=1.6in]{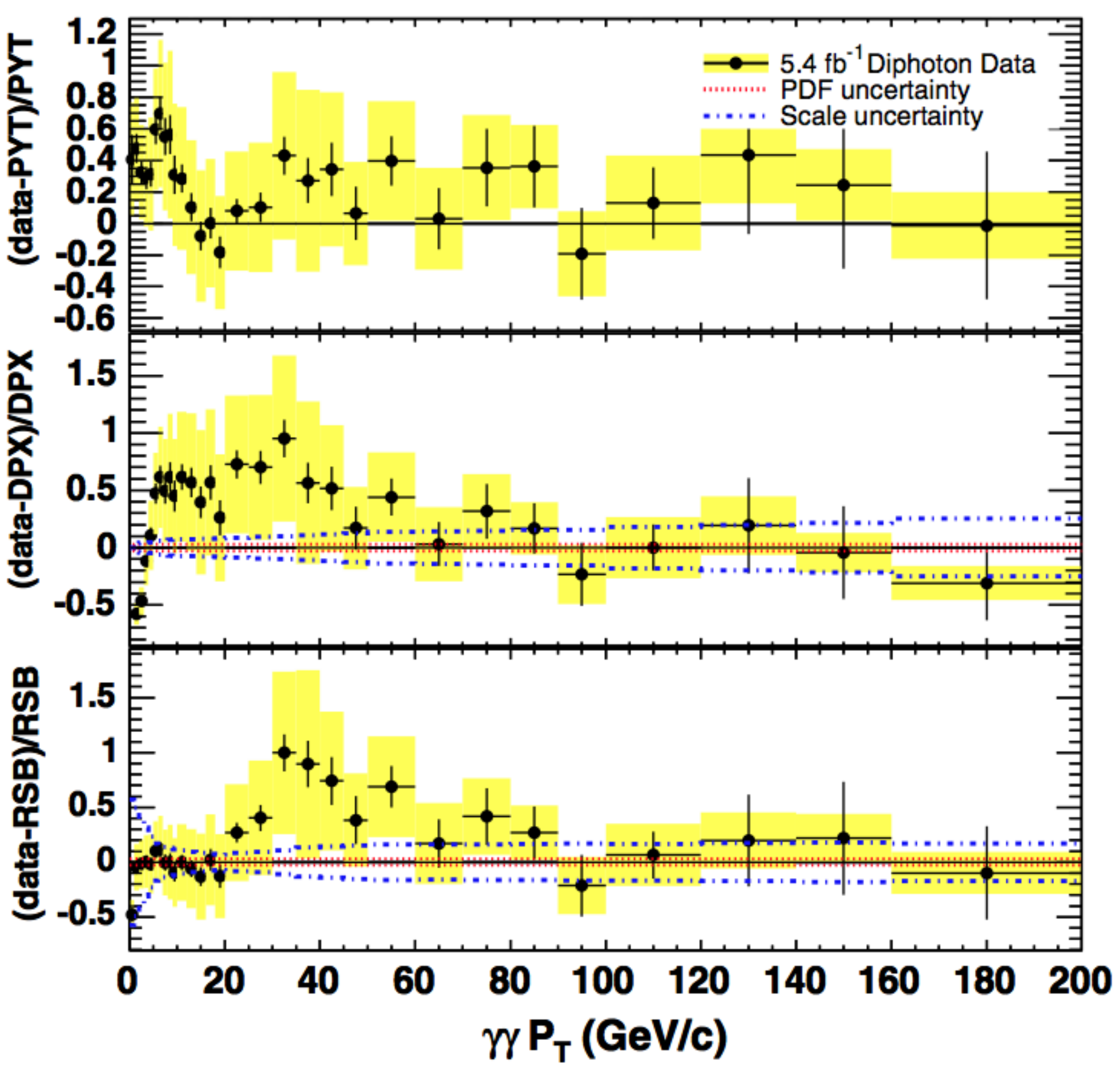}
\includegraphics[width=1.6in]{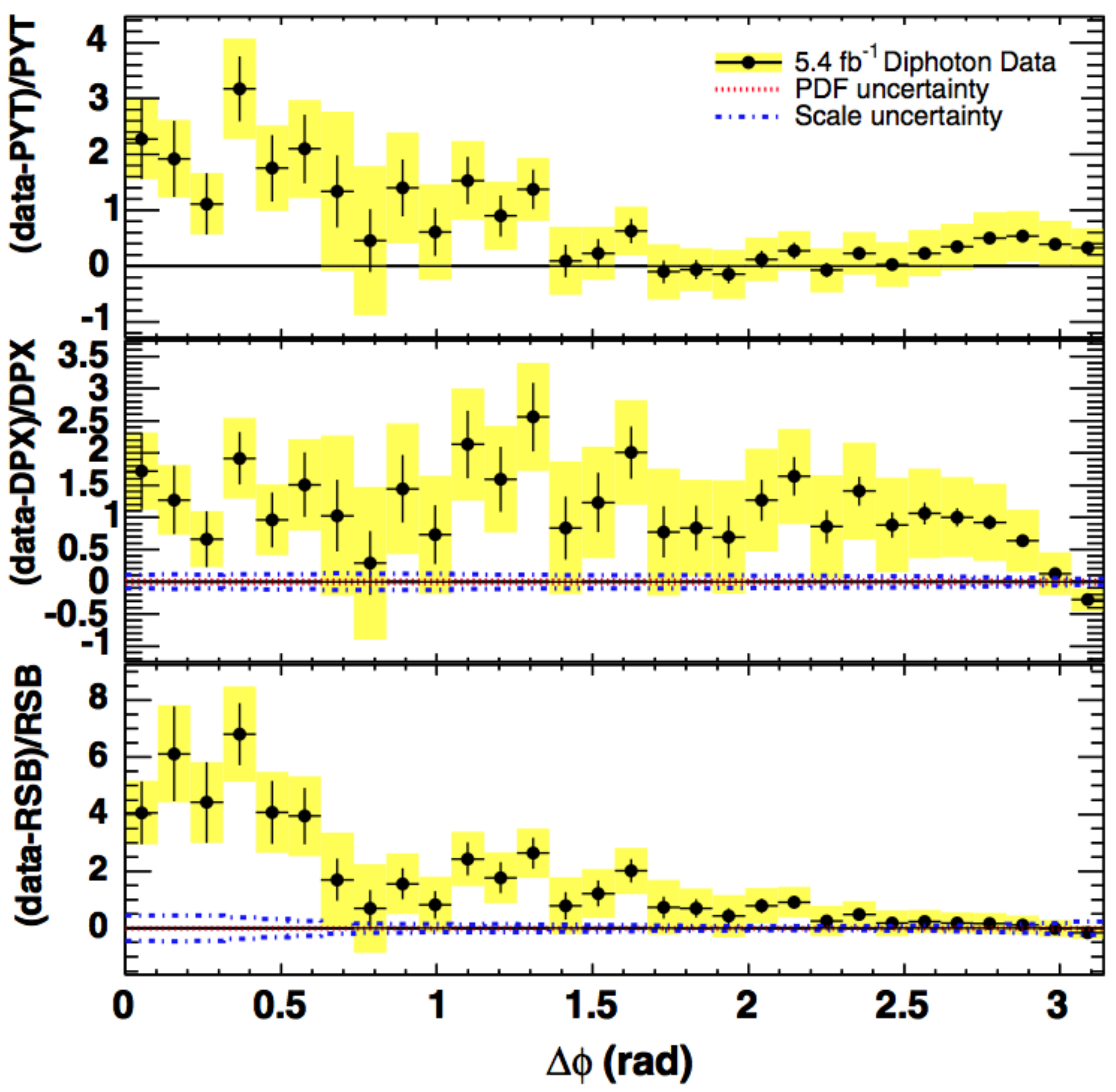}
\end{center}
\vspace*{-1ex}
\caption{The measured differential diphoton production cross sections at CDF as a function of (left) $M_{\gamma\gamma}$, (center) $p_T^{\gamma\gamma}$, and (right) $\Delta\phi_{\gamma\gamma}$.  Top: the absolute cross section values.  Bottom: the relative deviations of the data from predictions using \textsc{resbos}, \textsc{diphox}, and \textsc{pythia}.
\label{fig:cdf}}
\end{figure}

The results of the analyses are shown in Figures~\ref{fig:d0} and \ref{fig:cdf} for three kinematic variables: the diphoton invariant mass $M_{\gamma\gamma}$, the transverse momentum of the diphoton system $p_T^{\gamma\gamma}$, and the azimuthal angle between the photons $\Delta\phi_{\gamma\gamma}$.  All three calculations studied (\textsc{pythia}, \textsc{diphox}, and \textsc{resbos}) reproduce the main features of the data within their known limitations, but none of them describes all aspects of the data.  In the D0 analysis, \textsc{resbos} shows the best agreement with data, although systematic discrepancies are observed at low $M_{\gamma\gamma}$, high $p_T^{\gamma\gamma}$, and low $\Delta\phi_{\gamma\gamma}$.  The results from CDF are similar, and it is observed that the inclusion of photon radiation in the initial and final states significantly improves the \textsc{pythia} parton shower calculation.  The comparison between data and theory clearly indicates the necessity of including higher-order corrections beyond NLO, as well as the resummation of soft and collinear initial-state gluons to all orders.

\section{Angular Decorrelations in $\gamma$ + 2 and $\gamma$ + 3 Jet Events at D0}

The D0 collaboration uses data corresponding to 1.0 \invfb of integrated luminosity to measure differential cross sections versus azimuthal angles in $\gamma$ + 2 and $\gamma$ + 3 jet events.\cite{dzaz}  The purpose of this analysis is (1) to better understand non-perturbative QCD and to improve multiple parton interaction (MPI) models, (2) to learn new and complementary information about the spacial distribution of partons within the proton and correlations between them, and (3) to obtain better background estimates for other analyses such as Higgs boson searches.

\begin{figure}[t!]
\begin{center}
\includegraphics[width=1.3in]{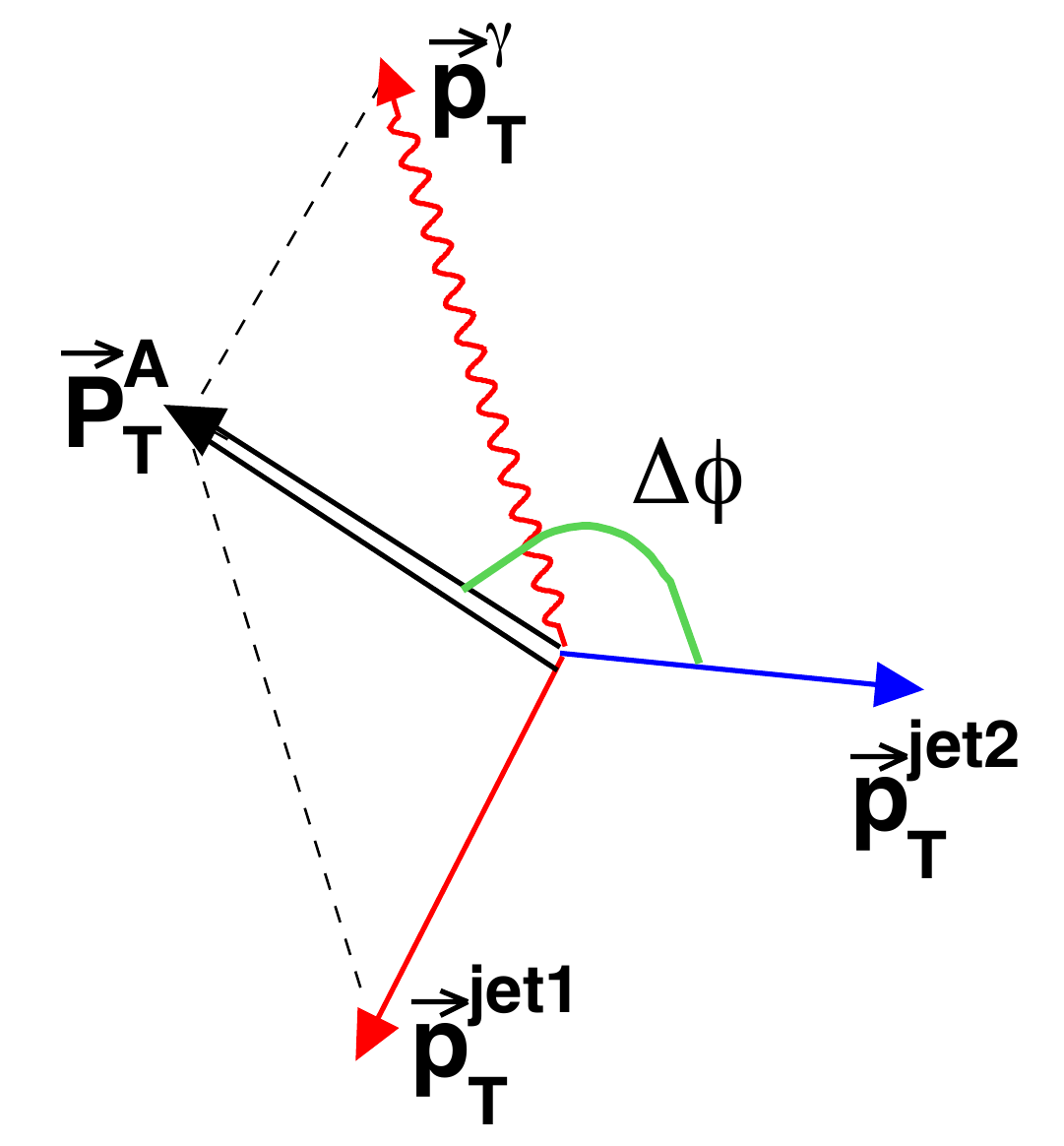}\hspace{0.2in}
\includegraphics[width=1.3in]{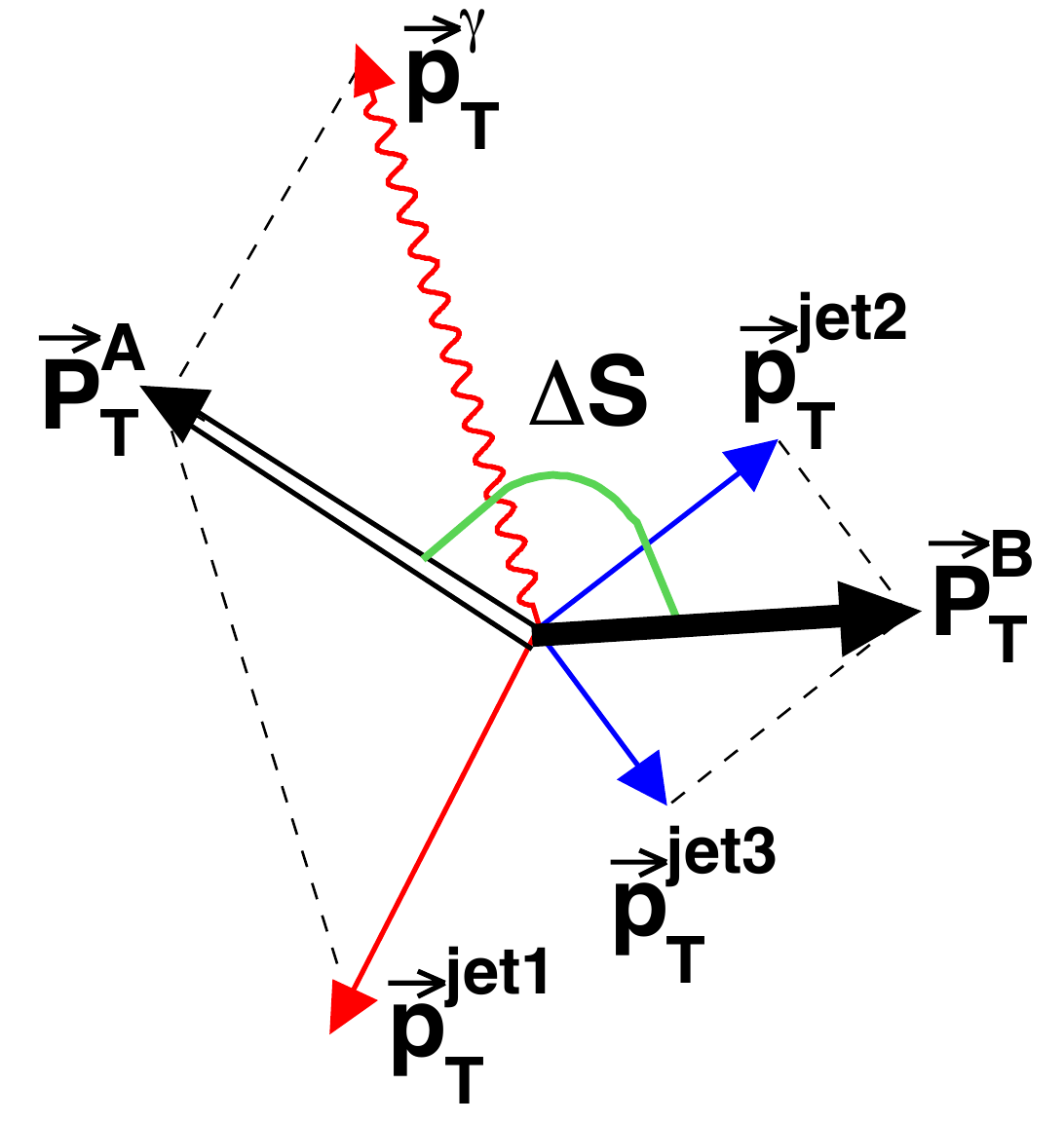}\hspace{0.2in}
\includegraphics[width=1.1in]{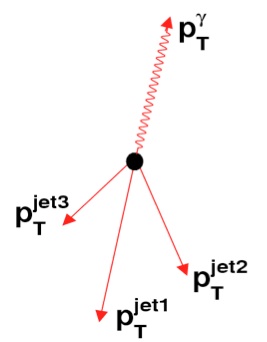}\hspace{0.2in}
\includegraphics[width=1.1in]{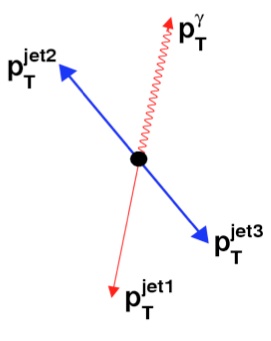}
\end{center}
\hfill {\fns (a)} \hspace{1.3in} {\fns (b)} \hspace{1.2in} {\fns (c)} \hspace{1.05in} {\fns (d)} \hfill \hfill \\*[-3ex]
\caption{(a) Diagram illustrating the definition of $\Delta\phi$ as the azimuthal angle between the $p_T$ vector of the $\gamma$ + leading jet system and the $p_T$ vector of jet2 in $\gamma$ + 2 jet events, (b) Diagram illustrating the definition of $\Delta S$ as the azimuthal angle between the $p_T$ vectors of the $\gamma$ + leading jet system and the jet2 + jet3 system in $\gamma$ + 3 jet events, (c) Single parton-parton (SP) interactions yield $\Delta\phi$ and $\Delta S$ distributions that are peaked at $\pi$, (d) Double parton (DP) interactions yield $\Delta\phi$ and $\Delta S$ distributions that are flat because there is no correlation between the separate parton-parton interactions.
\label{fig:dzazintro}}
\end{figure}

In this analysis, two kinematic quantities ($\Delta\phi$ and $\Delta S$) are defined that distinguish between single parton-parton (SP) interactions, in which the photon and all jets originate from the same hard scattering process with gluon bremsstrahlung in the initial or final state, and double parton (DP) interactions, in which two independent parton-parton interactions produce the photon + jets final state (see Figure~\ref{fig:dzazintro}).

\begin{figure}[b!]
\begin{center}
\includegraphics[width=1.54in]{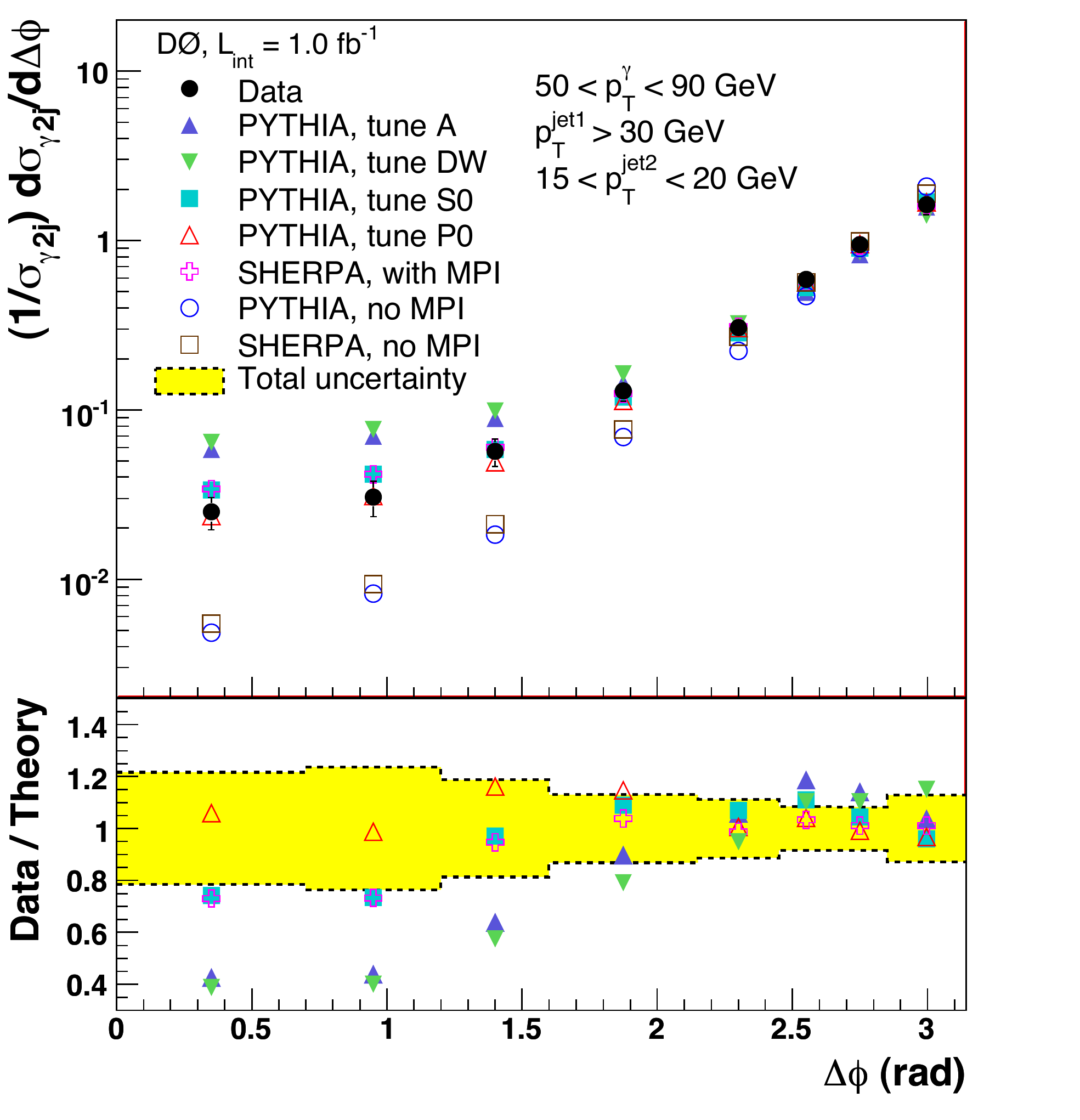}
\includegraphics[width=1.54in]{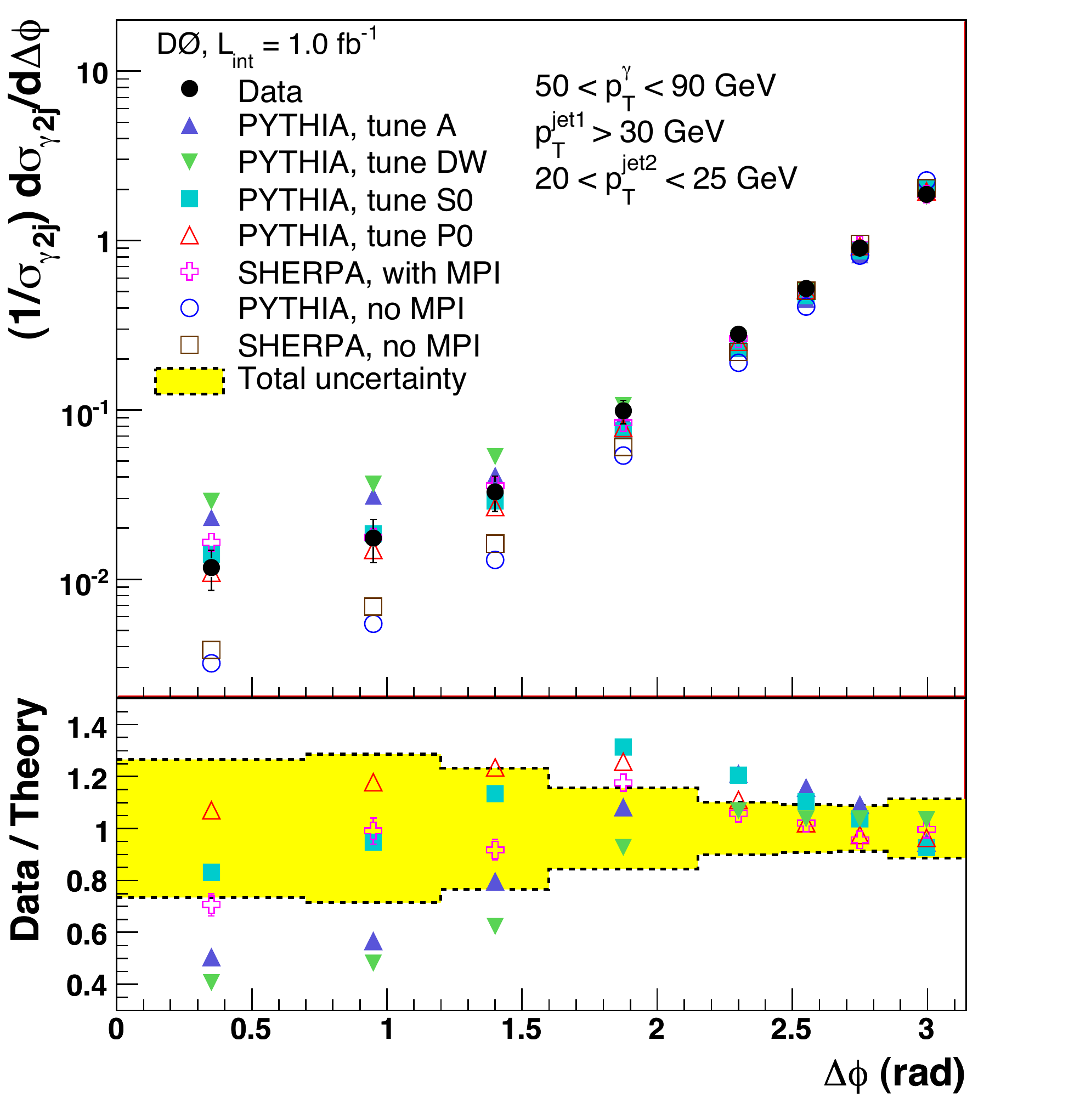}
\includegraphics[width=1.54in]{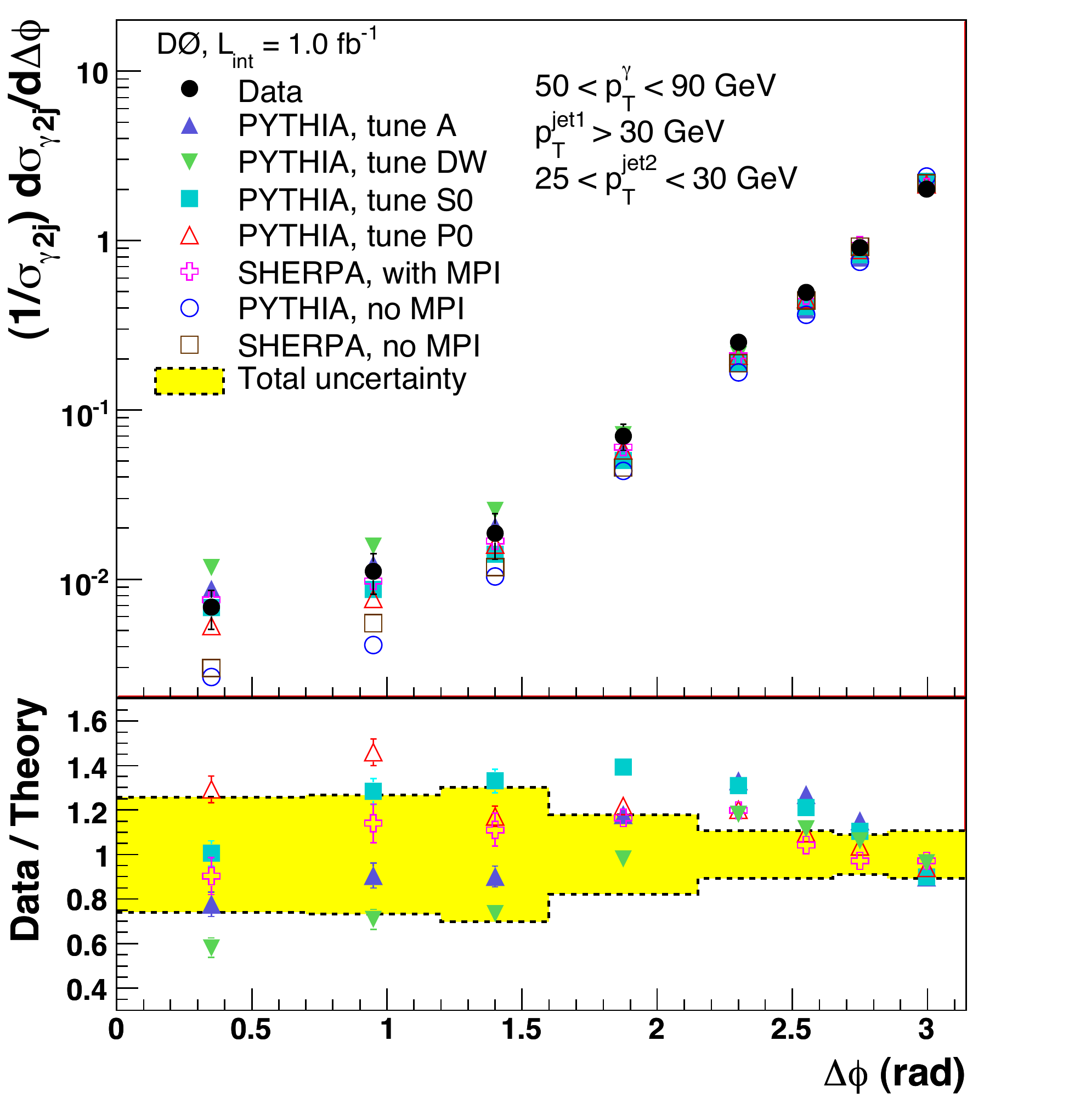}
\includegraphics[width=1.54in]{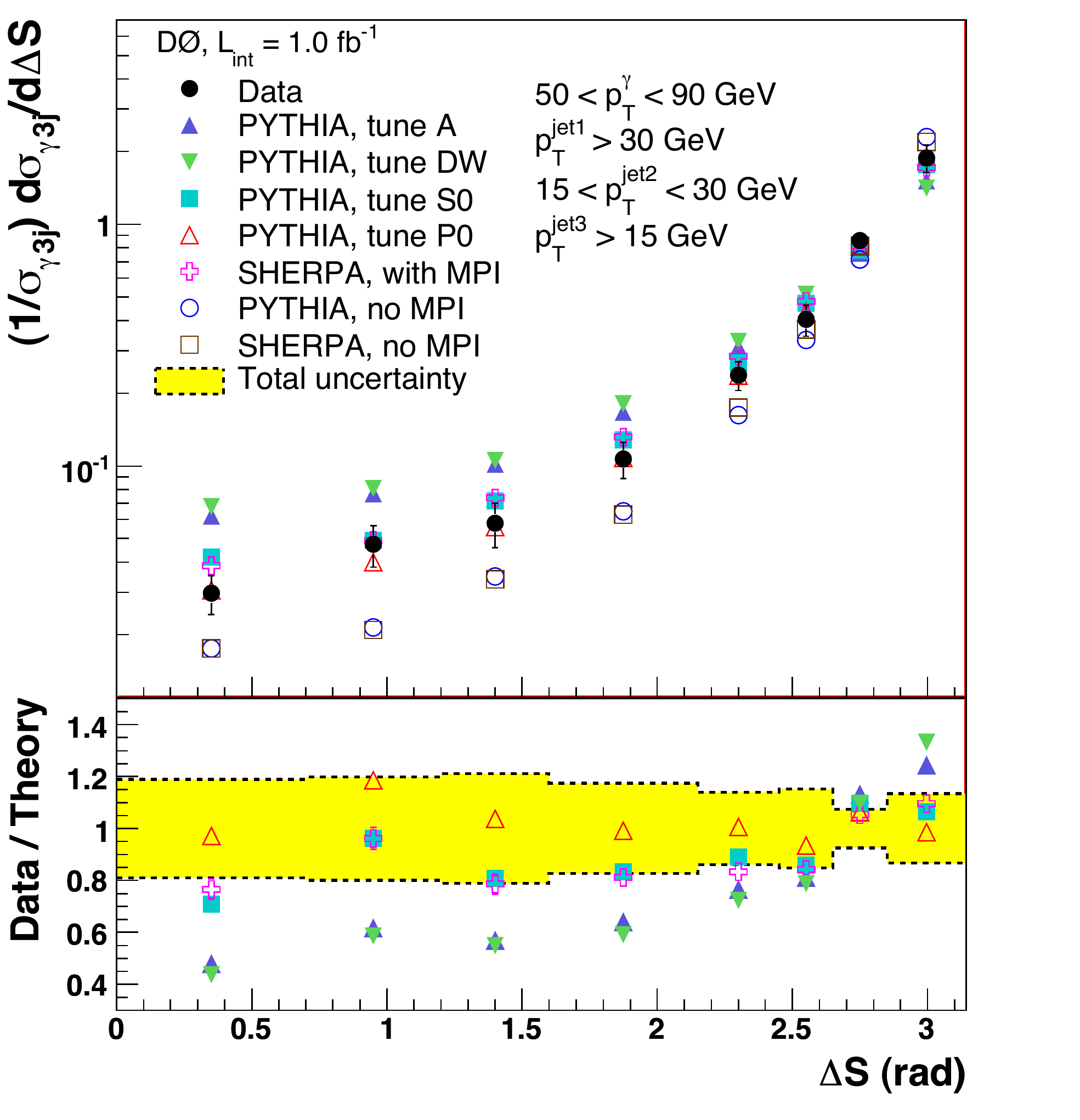}
\end{center}
\vspace*{-1ex}
\hfill {\fns (a)} \hspace{1.3in} {\fns (b)} \hspace{1.3in} {\fns (c)} \hspace{1.3in} {\fns (d)} \hfill \hfill \\*[-3ex]
\caption{(a)--(c) The measured normalized differential cross section in $\gamma$ + 2 jet events, $(1/\sigma_{\gamma 2j})d\sigma_{\gamma 2j}/d\Delta \phi$, compared to MC models for the ranges (a) 15 $< p_T^\mathrm{jet2} <$ 20 GeV, (b) 20 $< p_T^\mathrm{jet2} <$ 25 GeV, and (c) 25 $< p_T^\mathrm{jet2} <$ 30 GeV.  The ratio of data over theory is also provided (only for models including MPI).  (d) The measured normalized differential cross section in $\gamma$ + 3 jet events, $(1/\sigma_{\gamma 3j})d\sigma_{\gamma 3j}/d\Delta S$, compared to MC models for the range 15 $< p_T^\mathrm{jet2} <$ 30 GeV.  The ratio of data over theory is also provided (only for models including MPI).
\label{fig:dzazresults}}
\end{figure}

The results are summarized in Figure~\ref{fig:dzazresults}, which shows (1) the normalized differential cross section versus $\Delta\phi$ in $\gamma$ + 2 jet events for three bins of $p_T^\mathrm{jet2}$, and (2) the normalized differential cross section versus $\Delta S$ in $\gamma$ + 3 jet events.  Comparisons to theoretical predictions using \textsc{pythia} and \textsc{sherpa} reveal that the predictions of SP models alone do not provide an adequate description of the data; additional DP models are required.  The new \textsc{pythia} MPI models with $p_T$-ordered showers are favored, as well as the default \textsc{sherpa} showers.

\section{Exclusive Diphoton Production at CDF}

\begin{figure}
\begin{center}
\includegraphics[width=1.1in]{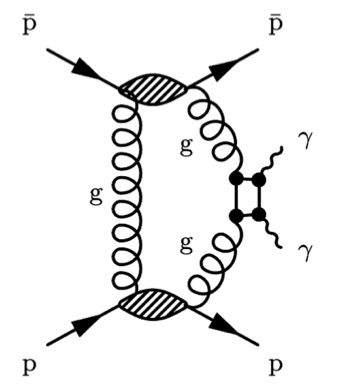}
\includegraphics[width=1.1in]{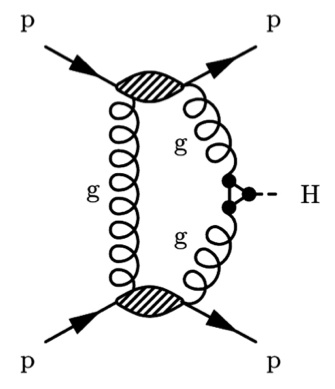}
\includegraphics[width=1.6in]{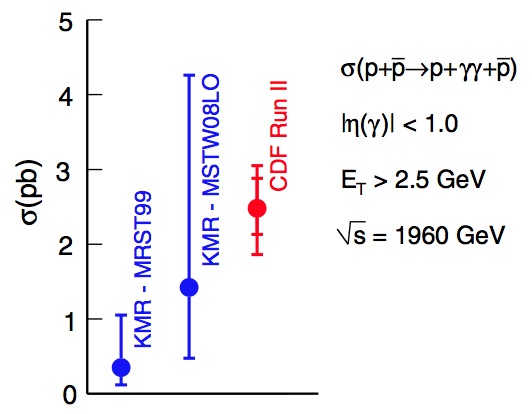}
\end{center}
\vspace*{-1ex}
\hfill {\fns (a)} \hspace{0.9in} {\fns (b)} \hspace{1.35in} {\fns (c)} \hfill \hfill \\*[-3ex]
\caption{(a) Leading-order diagram for central exclusive $\gamma\gamma$ production in \ppb collisions. (b) Leading-order diagram for central exclusive Higgs boson production in $pp$ collisions.  (c) Comparison of the measured cross section for exclusive $\gamma\gamma$ production in \ppb collisions at $\sqrt{s}$ = 1.96~TeV with theoretical predictions.
\label{fig:cdfexcl}}
\end{figure}

The CDF collaboration performed a search for exclusive $\gamma\gamma$ production via $\ppb \to p + \gamma\gamma + \overline{p}$ in data from 1.11~\invfb of integrated luminosity.\cite{cdfex}  This process is intrinsically interesting as a QCD process; moreover, it tests the theory of exclusive Higgs boson production in $pp$ collisions at the LHC.  Feynman diagrams of these processes are shown in Figure~\ref{fig:cdfexcl} (a) and (b).  Three features are evident in these events: (1) the proton and antiproton emerge intact with no hadrons produced, (2) the outgoing proton and antiproton have nearly the beam momentum ($p_T <$~1~GeV/c), and (3) rapidity gaps are located adjacent to the proton and antiproton.  The event selection requires two well reconstructed central ($|\eta| <$ 1.0) photons with $E_T >$~2.5~GeV and an absence of other activity in the detector.  Events with pileup are rejected.

After a careful treatment of background processes that produce an exclusive $\gamma\gamma$ final state (e.g. $q\overline{q} \to \gamma\gamma$), exclusive diphoton production was observed and the cross section for $\ppb \to p~+~\gamma\gamma~+~\overline{p}$ with $|\eta(\gamma)| <$~1.0 and $E_T(\gamma) >$~2.5~GeV was measured to be 2.48$^{+0.40}_{-0.35}$(stat)$^{+0.40}_{-0.51}$(syst) pb.  As shown in Figure~\ref{fig:cdfexcl}~(c), this cross section is in agreement with the only theoretical prediction, based on g + g $\to$ $\gamma$ + $\gamma$, with another gluon exchanged to cancel the color and with the $p$ and $\overline{p}$ emerging intact.  If a Higgs boson exists, it should be produced by the same mechanism and the cross sections are related.

\section*{Acknowledgments}
We thank the Fermilab staff, the technical staffs of the participating institutions, and the funding agencies for their vital contributions.

\end{document}